\documentclass[twocolumn,showpacs,preprintnumbers,amsmath,amssymb]{revtex4}
\usepackage{graphicx,bm}

\def\be{\begin{equation}}
\def\ee{\end{equation}}
\def\bea{\begin{eqnarray}}
\def\eea{\end{eqnarray}}
\def\nnb{\nonumber}
\def\bbuildrel#1_#2^#3{\mathrel{\mathop{\kern 0pt#1}\limits_{#2}^{#3}}}
\def\slash#1{\setbox0=\hbox{$#1$}#1\hskip-\wd0\dimen0=5pt\advance
       \dimen0 by-\ht0\advance\dimen0 by\dp0\lower0.5\dimen0\hbox
         to\wd0{\hss\sl/\/\hss}}

\newcommand{\scs}{\scriptscriptstyle}
\newcommand{\f}{\frac}

\begin{document}
\preprint{IFT-2/2008}
\title{Anomalous {\boldmath $Wtb$} coupling effects in the weak radiative {\boldmath $B$}-meson decay}
\author{Bohdan Grzadkowski and Miko{\l}aj Misiak}
\affiliation{Institute of Theoretical Physics, University of Warsaw, PL-00-681 Warsaw, Poland}
\affiliation{Theoretical Physics Division, CERN, CH-1211 Geneva 23, Switzerland}
\date{February 7, 2008}
\begin{abstract}
We study the effect of anomalous $Wtb$ couplings on the $\bar{B}\to
X_s \gamma$ branching ratio.  The considered couplings are introduced
as parts of gauge-invariant dimension-six operators that are built out
of the Standard Model fields only. One-loop contributions from the
charged-current vertices are assumed to be of the same order as the
tree-level flavour-changing neutral current ones. Bounds on the
corresponding Wilson coefficients are derived.
\end{abstract}
\maketitle
\section{Introduction \label{sec:intro}}

The large $\bar t t$ production cross section at the LHC is expected to provide
an opportunity to study $Wtb$ interactions with high accuracy (see, e.g.,
\cite{AguilarSaavedra:2007rs,Najafabadi:2008pb}). When performing such
studies, one should take into account constraints from the flavour changing
neutral current processes where loops involving top quarks play a crucial
role.  In particular, the inclusive decay $\bar{B} \to X_s \gamma$ provides
stringent bounds on the structure of $Wtb$ vertices.

In the present paper, we calculate contributions to the $\bar{B} \to X_s
\gamma$ branching ratio from one-loop diagrams involving several dimension-six
effective operators that give rise to non-standard $Wtb$ interactions.  We
work in the framework of an effective theory that is given by the Lagrangian
\be
{\cal L} = {\cal L}_{\rm SM} 
+ \f{1}{\Lambda  } \sum_i C_i^{(5)} Q_i^{(5)} 
+ \f{1}{\Lambda^2} \sum_i C_i^{(6)} Q_i^{(6)} 
+ {\cal O}\left(\f{1}{\Lambda^3}\right), \label{Leff}
\ee 
where ${\cal L}_{\rm SM}$ is the Standard Model (SM) Lagrangian, while
$Q_i^{(n)}$ denote dimension-$n$ operators that are invariant under the SM
gauge symmetries and are built out of the SM fields.  Such an approach is
appropriate for any SM extension where all the new particles are heavy
($M_{\rm new} \sim \Lambda \gg m_t$). So long as only processes at momentum
scales $\mu \ll \Lambda$ are considered, the heavy particles can be decoupled
\cite{Appelquist:1974tg}, which leads to the effective theory (\ref{Leff}).
Recent analyses of the top-quark anomalous couplings in the same framework
can be found, e.g., in Refs.~\cite{Fox:2007in,Cao:2007ea}.  

A complete classification of the operators $Q_i^{(5)}$ and $Q_i^{(6)}$ has
been given in Ref.~\cite{Buchmuller:1985jz}. Since $Q_i^{(5)}$ involve no
quark fields, we ignore them from now on, and skip the superscripts ``(6)'' at
the dimension-six operators and their Wilson coefficients $C_i$. Here, we
restrict our considerations to the following dimension-six operators that
generate anomalous $Wtb$ couplings:
\bea 
Q_{RR} &=& \bar{t}_R \gamma^\mu b_R \left( \widetilde{\phi}^\dagger i D_{\mu} \phi\right) + {\rm H.c.},\nnb\\
Q_{LL} &=& \bar{q}_L \tau^a \gamma^\mu q_L \left( \phi^\dagger \tau^a i D_{\mu} \phi\right) 
           -\bar{q}_L \gamma^\mu q_L \left( \phi^\dagger i D_{\mu} \phi\right)\nnb\\ 
        &+& {\rm H.c.},\nnb
\eea
\bea
Q_{LRt} &=& \bar{q}_L \sigma^{\mu\nu} \tau^a t_R \widetilde{\phi} W^a_{\mu\nu}
             + {\rm H.c.},\nnb\\
Q_{LRb} &=& \bar{q\,}_{\!\!L}' \sigma^{\mu\nu} \tau^a b_R \phi W^a_{\mu\nu} + {\rm H.c.},
\label{ops}
\eea 
where $\phi$ denotes the Higgs doublet,~ $\widetilde{\phi} = i\tau^2\phi^*$,
\bea
q_L &=&  \left( t_L,~ V_{tb} b_L + V_{ts} s_L + V_{td} d_L \right),\nnb\\
{q\,}_{\!\!L}' &=& \left( V^*_{tb} t_L + V^*_{cb} c_L + V^*_{ub} u_L,~ b_L \right),
\label{qfields}
\eea
and $V$ stands for the Cabibbo-Kobayashi-Maskawa (CKM) matrix. The $Wtb$ interaction vertex 
\bea
\includegraphics[width=25mm,angle=0]{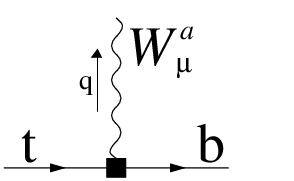}&&\nnb\\[-12mm]
&& \hspace{-5mm} = -\f{ig_{\rm w}}{\sqrt{2}} \left[ \gamma_\mu (v_L P_L + v_R P_R ) \right. \nnb\\
&& \left. + \f{i \sigma_{\mu\nu} q^\nu}{M_W} (g_L P_L + g_R P_R ) \right] \label{vertex}
\eea
with $P_{L,R} = \f{1}{2}(1\mp\gamma_5)$ is found by combining the usual SM
interaction with the extra contributions that are obtained by
setting the Higgs field in Eq.~(\ref{ops}) to its vacuum expectation value.

Our operators (\ref{ops}) have been adjusted to generate the vertex
(\ref{vertex}) in a gauge-invariant manner, without introducing extra
sources of CP-violation or tree-level Flavour Changing Neutral Current (FCNC)
interactions. The absence of tree-level FCNC in $Q_{RR}$, $Q_{LRt}$ and
$Q_{LRb}$ is transparent. Verifying that $Q_{LL}$ is also free of tree-level
FCNC requires a short calculation that is most conveniently performed in the
unitary gauge when the pseudogoldstone components of $\phi$ are absent. The
relative sign between the two parts of $Q_{LL}$ causes cancellation of FCNC
couplings like $\bar{s}_L \gamma_{\mu} b_L Z^{\mu}$. We wish to avoid such
couplings here because they would contribute at the tree level to the observed
decay $\bar{B} \to X_s l^+ l^-$.

%
%

Since our goal is testing anomalous couplings of the top quark without
affecting top-less physics, the flavour structure of $Q_{RR}$, $Q_{LL}$ and
$Q_{LRt}$ has been arranged in such a way that all the charged-current
interactions in these operators involve the top. The operator $Q_{LRb}$ does
not fulfill this requirement. It contains some $Wcb$ and $Wub$ vertices, too.
Using $q_L$ instead of $q'_L$ in this operator would cause problems with
tree-level FCNC. Thus, our final $\bar{B} \to X_s \gamma$ results are going to
receive contributions not only from the $Wtb$ vertex (\ref{vertex}) but also
from the $Wcb$ and $Wub$ parts of $Q_{LRb}$, from the $Wts$ and $\bar{t} t
\gamma$ parts of $Q_{LRt}$ (see Fig.~\ref{fig:ttg} in the next section), 
  from the $Wts$ part of $Q_{LL}$, and from flavour-off-diagonal
  field renormalization in the $\bar bb\gamma$ part of $Q_{LRb}$. The
appearance of non-$Wtb$ interactions is an unavoidable consequence of
introducing the anomalous $Wtb$ ones in a gauge-invariant manner.

It is important to realize that the particular flavour structure of our
operators should actually be set in the interaction basis, before the Yukawa
matrix diagonalization. This can be achieved by a proper alignment of the
Yukawa matrices and the couplings at the dimension-six operators. Here, we
shall not deliberate whether such an alignment can be natural in some
particular model. Our approach is purely phenomenological. Since the anomalous
$Wtb$ couplings (\ref{vertex}) are going to be investigated at the LHC, we
would like to know the current bounds on them from $\bar{B} \to X_s \gamma$,
assuming a particular embedding (\ref{ops}) of these couplings into
higher-dimensional operators.

The dimensionless couplings $v_{L,R}$ and $g_{L,R}$ in Eq.~(\ref{vertex})
are related to the Wilson coefficients $C_i$ as follows:
\bea
v_L &=& V^*_{tb} + \f{C_{LL} V^*_{tb}}{\sqrt{2} G_F \Lambda^2}, \hspace{1cm}
v_R = \f{C_{RR}}{2 \sqrt{2} G_F \Lambda^2}, \nnb\\[2mm]
g_L &=& \f{C_{LRb} V^*_{tb}}{G_F \Lambda^2}, \hspace{23mm}
g_R = \f{C_{LRt} V^*_{tb}}{G_F \Lambda^2}, \label{def.vvgg}
\eea
where $G_F = 2^{-\f{5}{2}} M_W^{-2} g_{\rm w}^2$ is the Fermi constant. The
coefficients $C_i$ are real, which follows from the fact that the operators in
Eq.~(\ref{ops}) are self-conjugate. Note that all these operators become
CP-even in the limit when the CKM matrix in Eq.~(\ref{qfields}) becomes real.

Constraints from ${\cal B}(\bar{B} \to X_s \gamma)$ on anomalous $Wtb$
couplings have already been studied in
Refs.~\cite{Larios:1999au,Burdman:1999fw}. However, those analyses were
restricted to the couplings $v_{L,R}$ in Eq.~(\ref{vertex}).  Moreover, our
results for the branching ratio dependence on $v_L$ are substantially
different, because an operator containing the $Wcb$ and $Wub$ vertices was
effectively used there instead of $Q_{LL}$.  

Our paper is organized as follows. In Sec.~\ref{sec:match}, we describe the
matching computation for passing from the effective theory (\ref{Leff}) to
another low-energy effective theory where the top quark and the electroweak
gauge bosons are already decoupled. In Sec.~\ref{sec:num}, a numerical
expression for the $\bar{B} \to X_s \gamma$ branching ratio as a function of
$v_{L,R}$ and $g_{L,R}$ is presented, and bounds on these parameters are
discussed. We conclude in Sec.~\ref{sec:concl}.

\section{Matching \label{sec:match}}

In the decay $\bar{B} \to X_s \gamma$, all the external momenta are much
smaller than $M_W$. Consequently, it is convenient to decouple the top quark
and the electroweak gauge bosons at the scale $\mu_0 \sim m_t, M_W$. At this
scale, we match the effective theory (\ref{Leff}) with another one, whose
Lagrangian has precisely the same form as in the SM case \cite{Grinstein:vj}
\be \label{Leff2}
{\cal L_{\rm eff}} =  {\cal L}_{\scs {\rm QCD} \times {\rm QED}}(u,d,s,c,b)
~+~ \f{4 G_F}{\sqrt{2}} V_{ts}^* V_{tb} \sum_{i=1}^8 C_i(\mu) Q_i,
\ee
where $Q_1$,~...,~$Q_6$ are four-quark operators, and 
\bea
Q_7 &=& \f{e m_b}{16 \pi^2} \bar{s}_L \sigma^{\mu \nu} b_R F_{\mu \nu},\nnb\\
Q_8 &=& \f{g_{\rm s} m_b}{16 \pi^2} \bar{s}_L \sigma^{\mu \nu} T^a b_R G^a_{\mu \nu}.
\eea
The presence of non-SM terms in Eq.~(\ref{Leff}) causes deviations of
$C_i(\mu_0)$ in Eq.~(\ref{Leff2}) from their SM values
\be \label{def.deltaC}
C_i(\mu_0) = C^{\rm SM}_i(\mu_0) + \Delta C_i(\mu_0).
\ee
So long as $v_{L,R}$ and $g_{L,R}$ are treated as quantities of zeroth order
in the expansion in $g_{\rm w}$ and $g_{\rm s}$, the deviations $\Delta
C_7(\mu_0)$ and $\Delta C_8(\mu_0)$ are also of zeroth order, similarly to
$C^{\rm SM}_7(\mu_0)$ and $C^{\rm SM}_8(\mu_0)$.  On the other hand, extra
contributions to the Wilson coefficients of the four-quark operators
$Q_1$,~...,~$Q_6$ arise only at higher orders in $g_{\rm w}$ or $g_{\rm s}$, 
and will be neglected here.

Because of ultraviolet renormalization, it would be inconsistent to assume that no
other operators but $Q_{RR}$,~...,~$Q_{LRb}$ (\ref{ops}) are present in the
dimension-six part of the Lagrangian (\ref{Leff}).  Instead, we shall make a
weaker assumption, namely that the $\overline{\rm MS}$-renormalized Wilson
coefficients of all the other relevant operators in Eq.~(\ref{Leff}) at scales
of order $\mu_0$ satisfy
\be
\f{C_{\rm other}(\mu \sim \mu_0)}{G_F\Lambda^2} \sim {\cal O}(g^n_w),
~~~ n \geq 2.
\ee
Under such an assumption, only tree-level $b \to s \gamma$ and $b \to s g$
diagrams with insertions of such operators must be included in our
leading-order calculation of $\Delta C_7(\mu_0)$ and $\Delta C_8(\mu_0)$.
Denoting such ``primordial'' tree-level contributions by $C_7^{(p)}(\mu_0)$
and $C_8^{(p)}(\mu_0)$, we can express $\Delta C_{7,8}(\mu_0)$ as follows 
\bea \label{def.f}
\Delta C_i(\mu_0) &=& C_i^{(p)}(\mu_0) + \f{1}{V^*_{tb}} \left[ \delta v_L\, f_i^{v_L}(x) 
+ v_R \f{m_t}{m_b} f_i^{v_R}(x) 
\right.\nnb\\[2mm]
&+& \left. g_L \f{M_W}{m_b} f_i^{g_L}(x) + g_R \f{m_t}{M_W} f_i^{g_R}(x) \right],
\eea
where $x = m_t^2/M_W^2$ and $\delta v_L = v_L - V^*_{tb}$. It is understood
that the Wilson coefficients in the definitions (\ref{def.vvgg}) of $v_{L,R}$
and $g_{L,R}$ are $\overline{\rm MS}$-renormalized at the matching scale
$\mu_0$.

The functions $f_{7,8}^{v_{L,R}}(x)$ and $f_{8}^{g_{L,R}}(x)$ originate from
ultraviolet-finite diagrams, and depend on $x$ only. However, divergent
diagrams occur in the calculation of $f_7^{g_{L,R}}(x)$. Consequently,
logarithms $\;\ln\f{\mu_0}{M_W}$ are present in these functions.  They remain
after applying the $\overline{\rm MS}$ prescription for absorbing the
divergences into the operators in Eq.~(\ref{Leff}) that generate
$C_i^{(p)}(\mu_0)$.  Several operators can serve as the corresponding
counterterms --- see section 4.8 of Ref.~\cite{Buchmuller:1985jz}.  Our
final results for $f_i^{g_{L,R}}(x)$ can be (and are) found without making any
particular choice for the structure of these operators.

In Eq.~(\ref{def.f}) and everywhere in the following, non-linear terms in
$v_{L,R}$ and $g_{L,R}$ have been neglected. Including them in a consistent
manner would require extending the operator basis (\ref{ops}) to operators of
dimension higher than 6.  Consequently, our calculation is valid only for
$v_{L,R}, g_{L,R} \ll 1$, even though these quantities are formally treated as
being zeroth order in $g_{\rm w}$.

The functions $f_i^{v_L}(x)$ and $f_i^{v_R}(x)$ can be found without
performing any new Feynman diagram computation. A brief inspection into the
structure of $Q_{LL}$ and $Q_{RR}$ (most conveniently in the unitary gauge)
reveals that all the relevant Feynman diagrams are identical to those that
have already occurred either in the SM or in the LR-model \cite{Cho:1993zb}
analyses of $b \to s \gamma$. Explicitly (see Eqs.~(6) and (11) of
Ref.~\cite{Bobeth:1999mk} as well as Eqs.~(3.2) and (4.6) of
Ref.~\cite{Cho:1993zb}):
\bea
f_7^{v_L}(x) &=& \f{3x^3-2x^2}{2(x-1)^4} \ln x + \f{22x^3-153x^2+159x-46}{36(x-1)^3},\nnb\\[3mm]
f_7^{v_R}(x) &=& \f{-3x^2+2x}{2(x-1)^3} \ln x + \f{-5x^2+31x-20}{12 (x-1)^2}, \nnb\\[3mm]
f_8^{v_L}(x) &=& \f{-3x^2}{2(x-1)^4} \ln x + \f{5x^3-9x^2+30x-8}{12(x-1)^3},\nnb\\[3mm]
f_8^{v_R}(x) &=& \f{3x}{2 (x-1)^3} \ln x -\f{x^2+x+4}{4(x-1)^2}. \label{fvv}
\eea 
\begin{figure}[t]
\begin{center}
\includegraphics[width=4cm,angle=0]{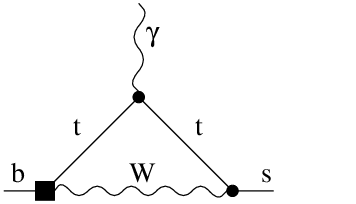}
\includegraphics[width=4cm,angle=0]{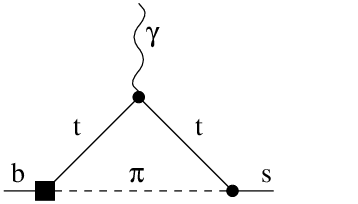}\\[3mm]
\includegraphics[width=4cm,angle=0]{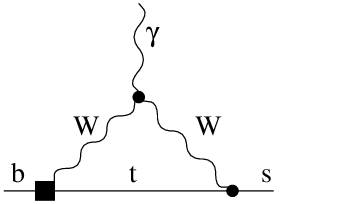}
\includegraphics[width=4cm,angle=0]{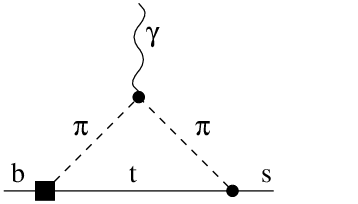}\\[3mm]
\includegraphics[width=4cm,angle=0]{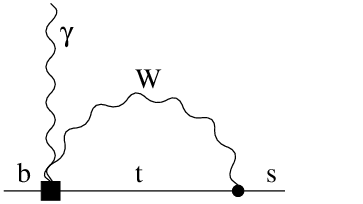}
\includegraphics[width=4cm,angle=0]{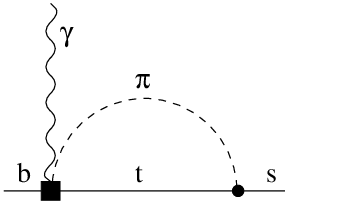}
\end{center}
\begin{center}
\caption{\sf Diagrams with non-SM $b\to t$ vertices that contribute to 
  $f_7^{g_{L,R}}(x)$. The pseudogoldstone boson is denoted by $\pi$. \label{fig:b2t}}
\end{center}
\vspace*{-4mm}
\end{figure}
\begin{figure}[t]
\begin{center}
\includegraphics[width=4cm,angle=0]{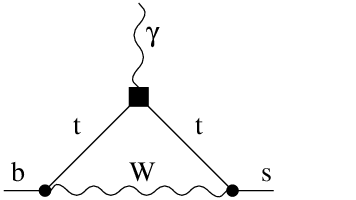}
\includegraphics[width=4cm,angle=0]{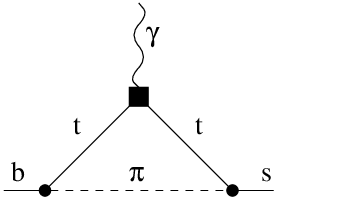}
\end{center}
\begin{center}
\caption{\sf Diagrams with non-SM $\bar{t}t\gamma$ vertices that contribute to 
  $f_7^{g_R}(x)$. \label{fig:ttg}}
\end{center}
\vspace*{-4mm}
\end{figure}

As far as $f_7^{g_{L,R}}(x)$ are concerned, our calculation of these
functions has been performed in the Feynman--'t~Hooft gauge. The relevant
Feynman diagrams with non-SM $b\to t$ vertices are shown in
Fig.~\ref{fig:b2t}.  In addition, analogous six diagrams with non-SM $t
\to s$ vertices and two diagrams with non-SM $\bar{t}t\gamma$ vertices 
(Fig.~\ref{fig:ttg}) occur in the case of $f_7^{g_R}$(x). In the case
of $f_7^{g_L}$(x), there are also diagrams where the intermediate $t$-quark
gets replaced by $u$ or $c$. The functions $f_8^{g_{L,R}}(x)$ have been
found by replacing the external photon by the gluon in the diagrams like
the ones in the first row of Fig.~\ref{fig:b2t}.

Our final results for $f_i^{g_{L,R}}(x)$ read:
\bea
f_7^{g_L}(x) &=& x \ln\f{\mu_0}{M_W} 
              + \f{-3x^4+16x^3-12x^2+2x}{6(x-1)^3} \ln x\nnb\\[2mm]
             &+& \f{6x^3-31x^2+19x}{12(x-1)^2},\nnb\\[3mm]
f_7^{g_R}(x) &=& -\f{1}{4} \ln\f{\mu_0}{M_W} 
              + \f{3x^4\!-\!12x^3\!-\!27x^2\!+\!32x\!-\!8}{24(x-1)^4} \ln x\nnb\\[2mm]
             &+& \f{-15x^3+97x^2-69x+11}{48(x-1)^3},\nnb\\[3mm]
f_8^{g_L}(x) &=& \f{2x^3-6x^2+x}{2(x-1)^3} \ln x
               + \f{x^2+5x}{4(x-1)^2},\nnb\\[3mm]
f_8^{g_R}(x) &=& \f{4x-1}{2(x-1)^4} \ln x 
              + \f{2x^2-9x+1}{4(x-1)^3}. \label{fgg}
\eea

The diagrams in Figs.~\ref{fig:b2t}~and~\ref{fig:ttg} correspond to an
off-shell calculation in the background-field gauge.  Calculating on shell
would bring some one-particle reducible diagrams into the game. Without the
background field method, one would need to include additional diagrams with
$W\gamma\pi$ couplings, where $\pi$ stands for the pseudogoldstone boson.  
We have actually performed the calculation using both methods, which has
served as a cross-check of the final result.

\section{Numerical Results \label{sec:num}}

Once the matching conditions are found, the calculation proceeds precisely as
in the SM case. For the purpose of this section, we shall assume that
$C_{7,8}^{(p)}(\mu_0)$ are real and neglect the imaginary part of $V_{tb}$.  The
$\bar{B}\to X_s\gamma$ branching ratio for arbitrary real values of $\Delta
C_{7,8}(\mu_0)$ reads \cite{Misiak:2006zs,Misiak:2006ab}
\bea
{\cal B} &\equiv& {\cal B}(\bar{B}\to X_s\gamma)_{E_\gamma > 1.6\,{\rm GeV}} \times 10^4 
= (3.15 \pm 0.23)\hspace{8mm}\nnb\\[2mm]
&-& 8.0\, \Delta C_7(\mu_0) - 1.9\,\Delta C_8(\mu_0)
+ {\cal O}\left[ (\Delta C_i)^2 \right]\!, \label{BRgen}
\eea
for the numerical inputs as specified in Appendix A of
Ref.~\cite{Misiak:2006ab}, in particular, $\mu_0 = 160\,$GeV.  Inserting our
results from Eqs.~(\ref{def.f})--(\ref{fgg}) into Eq.~(\ref{BRgen}), one finds
\bea 
{\cal B} &=& (3.15 \pm 0.23) - 8.2\, \delta v_L + 427\, v_R - 837\, g_L\nnb\\[2mm] 
&+& 1.9\, g_R - 8.0\, C_7^{(p)}(\mu_0) - 1.9\, C_8^{(p)}(\mu_0) \nnb\\[2mm]
&+& {\cal O}\left[ \left(\delta v_L,v_R,g_L,g_R,C_i^{(p)}\right)^2\right].\label{BRvvgg}
\eea

As the reader might have expected, the coefficients at $\delta v_L$ and $g_R$
are of the same order as the first (SM) term, while the coefficients at $v_R$
and $g_L$ are substantially larger. For $v_R$ and $g_L$, an
enhancement~\cite{Cho:1993zb,Fujikawa:1993zu} by $m_t/m_b$ takes place,
because the SM chiral suppression factor $m_b/M_W$ gets replaced by the
order-unity factor $m_t/M_W$. This was already evident in Eq.~(\ref{def.f}).

The negative coefficient at $\delta v_L$ in Eq.~(\ref{BRvvgg}) differs from
the positive one in Fig.~1 of Ref.~\cite{Burdman:1999fw} where the
leading-order (LO) expression for $C^{\rm SM}_7(\mu_0)$ was used instead of
our $f_7^{v_L}(x)$.  The two quantities have different signs due to an
additive constant in the relation
\be
C^{\rm SM}_7(\mu_0)_{LO} = \f{1}{2} f_7^{v_L}(x) - \f{23}{36}.
\ee
This constant originates from the SM loops where the top quark is replaced by
the light ones (up and charm).  No such loops are generated by our operator
$Q_{LL}$. The flavour structure of the operators in
Refs.~\cite{Larios:1999au,Burdman:1999fw} has not been specified in sufficient
detail.

The appearance of $\ln \mu_0/M_W$ in Eq.~(\ref{fgg}) implies that the
coefficients at $g_L$ and $g_R$ in Eq.~(\ref{BRvvgg}) strongly depend on
$\mu_0$. These coefficients are well approximated by 
$-171 - 969 \ln \mu_0/M_W$ and $-0.87 + 4.04 \ln \mu_0/M_W$, 
respectively.  Their $\mu_0$-dependence and the one of $C_i^{(p)}(\mu_0)$
should compensate each other in Eq.~(\ref{BRvvgg}), up to residual
higher-order effects.

\begin{table}[t]
\newlength{\minus} \settowidth{\minus}{$-$} \newcommand{\m}{\hspace{\minus}}
\begin{center}
\begin{tabular}{|c|c|c|c|c|c|c|} \hline
bound &$ \delta v_L $&$   v_R   $&$   g_L   $&$  g_R  $&$C_7^{(p)}$&$C_8^{(p)}$\\\hline
upper &$  \m0.03    $&$\m0.0025 $&$\m0.0004 $&$\m0.57 $&$    \m0.04         $&$     \m0.15        $\\
lower &$   -0.13    $&$ -0.0007 $&$ -0.0013 $&$ -0.15 $&$     -0.14         $&$      -0.56        $\\\hline
\end{tabular}
\caption{ The current $95\%\,$C.L. bounds from
  Eq.~(\ref{BRvvgg}) along the parameter axes for $\mu_0 = 160\,$GeV. \label{tab:bounds}}
\end{center}
\end{table}

Taking into account the current world average~\cite{Barberio:2007cr}
\be \label{bsg.hfag}
{\cal B} = 3.55 \pm 0.24{\;}^{+0.09}_{-0.10} \pm 0.03,
\ee
one finds that a thin layer in the six-dimensional space \linebreak 
$(\delta v_L, v_R, g_L, g_R, C_7^{(p)}(\mu_0), C_8^{(p)}(\mu_0))$ 
remains allowed by $b\to s\gamma$.  When a single parameter at a time is
varied around the origin (with the other ones turned off), quite narrow
$95\%\,$C.L. bounds are obtained.  They are listed in Table~\ref{tab:bounds}.
If several parameters are simultaneously turned on in a correlated manner,
their magnitudes are, in principle, not bound by $b\to s\gamma$ alone.
However, the larger they are, the tighter the necessary correlation is,
becoming questionable at some point.

It is interesting to compare Table~\ref{tab:bounds} with the sensitivity of
top quark decay observables to $v_R$, $g_L$ and $g_R$. The ATLAS study in
Ref.~\cite{AguilarSaavedra:2007rs} reveals that their measurements should allow to put
bounds on $g_R$ at the level of (a few)$\times 10^{-2}$, i.e. stronger than
the $\bar{B}\to X_s\gamma$ ones. On the other hand, the bounds they expect to
set on $v_R$ and $g_L$ are more than an order of magnitude weaker than those
in Table~\ref{tab:bounds}, which is due to the previously mentioned $m_t/m_b$
enhancement. 

As far as $\delta v_L$ is concerned, single top production measurements at the
Tevatron imply $\delta v_L = 0.3 \pm 0.2$ \cite{Abazov:2006gd}. Around an
order of magnitude smaller uncertainty is expected at the
LHC~\cite{O'Neil:2002ks}, which would definitely overcome the current
$\bar{B}\to X_s\gamma$ bounds.

\section{Conclusions \label{sec:concl}}

We have studied the effect of anomalous $Wtb$ couplings on the $\bar{B}\to X_s
\gamma$ branching ratio.  The couplings were introduced via gauge-invariant
dimension-six operators.  Our results for the branching ratio dependence on
$g_L$ and $g_R$ are new. In the case of $\delta v_L$, we have demonstrated the
necessity of precisely defining the flavour structure of the relevant
operators, which has not been previously done in sufficient detail.

The well-known $m_t/m_b$ enhancement~\cite{Cho:1993zb,Fujikawa:1993zu} implies
that the $\bar{B}\to X_s \gamma$ bounds on $v_R$ and $g_L$ are much stronger
than what one can possibly hope to obtain from studying the top quark
production and decay at the LHC. On the other hand, the future LHC bounds on
$\delta v_L$ and $g_R$ are expected to overcome the current $\bar{B}\to X_s
\gamma$ ones.

Considering other FCNC processes would increase the number of constraints but
also bring new FCNC operators with their Wilson coefficients into the game, so
long as the amplitudes undergo ultraviolet renormalization. Consequently, the
analysis would become more and more involved. Effects of $\delta v_L$ and
$v_R$ on $b\to s l^+ l^-$ have been discussed, e.g., in
Refs.~\cite{Burdman:1999fw,Lee:2002bm}.  These studies need to be updated in
view of the recent measurements, and extended to the case of $g_L$ and $g_R$.
The same refers to the $B\bar B$ mixing, for which (to our knowledge) no
dedicated calculation has been performed to date.

\begin{center} {\bf ACKNOWLEDGMENTS} \end{center}

We would like to thank G.~Burdman, N. Castro and M.~Mangano for motivating
remarks. We are grateful to Jure Drobnak, Svjetlana Fajfer and Jernej Kamenik
for letting us know prior to their publication~\cite{Drobnak:2011aa} that
contributions to $f_7^{g_L}$ in Eq.~(\ref{fgg}) from flavour-off-diagonal
field renormalization were missing in the previous version of this
article. This work has been supported in part by the Polish Ministry of
Science and Higher Education as a research project N~N202~006334 (in years
2008-11), by the EU-RTN Programme, contract no.~MRTN--CT-2006-035482,
FLAVIAnet, and by the Marie Curie Research Training Network HEPTOOLS, contract
no.~MRTN-CT-2006-035505.

\setlength {\baselineskip}{0.2in}
 
\end{document}